\documentclass[pre,floatfix,twocolumn, 10pt]{revtex4} 
\usepackage{graphicx}
\usepackage{amssymb}
\usepackage{natbib}
\usepackage{dcolumn}
\usepackage{bm}
\usepackage{color}
\usepackage[colorlinks=true, linkcolor=blue, citecolor=blue]{hyperref}
\usepackage{subfigure}
\usepackage{graphicx,psfrag,xspace}
\usepackage{color}
\usepackage{amssymb,amsfonts,amsmath}
\usepackage{setspace}

\begin{document}


\author{F. V. Pereyra Aponte and E. A. Jagla} 
\affiliation{Centro At\'omico Bariloche, Instituto Balseiro, 
Comisi\'on Nacional de Energ\'ia At\'omica\\ CNEA, CONICET, UNCUYO, Av.~Bustillo 9500 (R8402AGP), San Carlos de Bariloche,  Argentina}

\title{Global Oscillations in Depinning Models with Aging}
\begin{abstract} 

We propose a  model that extends the standard depinning paradigm by incorporating an aging mechanism into the local pinning force. This favors oscillations between a stuck state of large pinning, and a slipping state of smaller pinning.
We show that for mean field interactions between sites this mechanism
can lead to the appearance of ``king avalanches" and global instabilities, producing a global oscillatory stick-slip stress regime. We construct the phase diagram for this mean field case and identify regions of smooth dynamics, pure stick-slip, and bistability. Crucially, when considering  two-dimensional systems with short-range interactions we find that states of global stress oscillation persist, 
but in contrast to the mean field case, no system-size avalanches appear. Instead, we observe alternating temporal intervals of larger and lower avalanche activity that correlate with the stress oscillations.
\end{abstract}

\maketitle

\section{Introduction}

A wide range of out-of-equilibrium systems subject to a continuous input of energy respond through intermittent, abrupt bursts of activity, commonly referred to as avalanches. They include 
among other the cases of domain wall motion\citep{zapperi1998dynamics,durin2024earthquakelike},
crack propagation\citep{bonamy2008crackling,laurson2013evolution}, deformation of amorphous solids\citep{nicolas2018deformation} and fault dynamics\citep{BK67, de2016statistical}.
%
These avalanches are originated in a cascading effect in which a small instability in some part of the system triggers ulterior instabilities in other regions, leading in the end to an activity extending over large regions. The concept of avalanches has been generalized to a wide range of natural, biological, and social systems\cite{Tur99,LFM03,Bar05,Asc11,dALGM06,Mun18,Vig25}.
Avalanches are characterized by their size $S$, which is related to the amount of energy they dissipate. 
In the case of earthquakes, this size relates to the magnitude through a logarithmic relation, i.e., magnitude is linearly related to the logarithm of the energy dissipated by the earthquake. 
The size distribution of avalanches $P(S)$ is one of the most prominent characteristics of a given system displaying this kind of dynamic activity.
In many situations avalanches of all sizes appear during the dynamic evolution, and the  
size distribution of avalanches may follow a power law.
Theoretical models designed to describe the dynamics of these systems find indeed a power law distribution of avalanches. For instance, in the so called {\em depinning} paradigm \cite{kardar1998nonequilibrium,fisher1998collective,wiese2022theory}, the avalanche distribution $P(S)$ has a power law form with an exponent $\tau_0$, and a cutoff at some large size $S_{max}$, namely 
\begin{equation}
P(S)\sim S^{-\tau_0}f(S/S_{max}).
\end{equation}
The function $f(x)$ tends to 1 for $x\ll 1$, and to 0 for $x\gg 1$. In many cases, this function is found to be very well approximated by a simple exponential function: $f(x)\sim \exp(-x)$. The value of $S_{max}$ depends of some tuning parameter in the system. For instance, in the depinning implementation, the system may be driven by pulling on it with a spring with some stiffness $k_0$. In this case, $S_{max}$ depends on $k_0$ through some inverse power law, namely $S_{max}\sim k_0^{-\sigma}$ ($\sigma>0$), in such a way that $S_{max}$ diverges as $k_0$ goes to zero. It is said that the system becomes {\em critical} in this limit, with the avalanche distribution becoming a perfect power law, and therefore scale free, lacking a characteristic size for the avalanches.

The depinning paradigm has been the standard reference in many instances of systems with a broad distribution of avalanche sizes following a cutoff power law.  However, not always the avalanche distribution of a physical system is a cutoff power law.
A rather common situation is that of a system with a distribution of avalanches that is clearly separated in two distinctive parts. One is the ``normal" cutoff power law distribution, the other is a group of vary large avalanches that clearly stand out the normal power law\cite{Sor02,SoOu12}.
This coexistence has also been observed in numerical simulations, 
both in elastic interfaces\citep{ jagla2010mechanism, papanikolaou2012quasi, dansereau2019collective} and in amorphous solids\citep{salerno2013effect,karimi2017inertia, de2019collective, ruscher2021avalanches, martens2012spontaneous}. 

Avalanches that stand out of the cutoff power law may have a prominent effect in the full dynamics of the system under study.
These avalanches have been called {\em king} (or {\em dragon king}) avalanches \cite{Sor02,Sor09,SoOu12}, and are responsible for abrupt drops in the loading state (tipically measured by some internal stress) of the system. They can lead, in the case of a quasi-periodic appearance, to an oscillatory  behavior of the stress, with loading periods with very few, small avalanches, interrupted by kings that globally unload the system, resetting its accumulated stress.

A number of mechanisms have been proposed as possible origin of king avalanches. In general these mechanisms imply to consider a dynamical evolution that is not limited to the overdamped limit that is usually assumed in the depinning models. 
Such conditions arise, for example, in the presence of inertia\citep{salerno2013effect,karimi2017inertia, de2019collective, de2022scaling}, stress overshoot\cite{fisher1997statistics,schwarz2001depinning,schwarz2003depinning}, viscoelasticity \citep{marchetti2003driven,marchetti2005models,marchetti2006depinning,jagla2014viscoelastic}, internal relaxation processes \citep{jagla2010mechanism, papanikolaou2012quasi, martens2012spontaneous,dansereau2019collective}, or facilitation mechanisms \cite{dSBVM16}.

\begin{figure}
\includegraphics[width=8cm,clip=true]{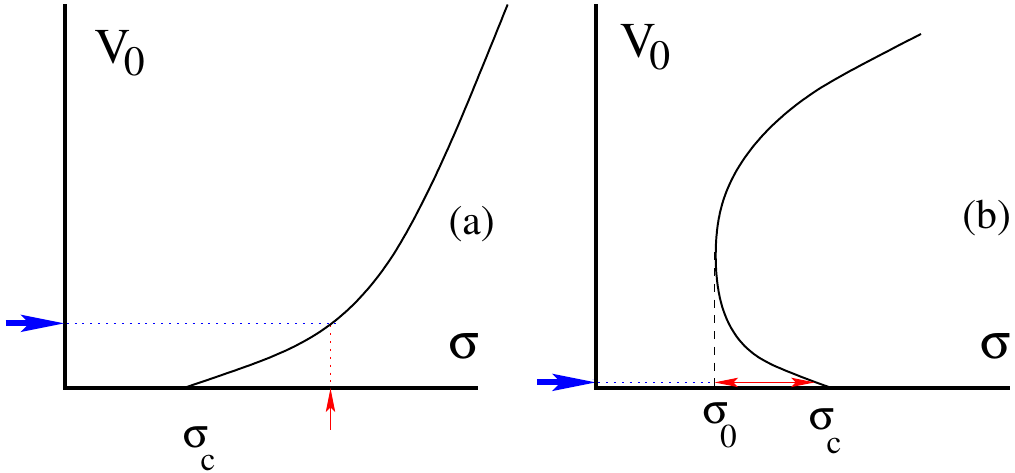}
\caption{Normal (a) and reentrant (b) flow curves. In the first case, driving at any constant velocity produces a unique (up to fluctuations) value of the stress. In the last case, when driving at a constant small velocity (thick blue arrow), the stress is expected to oscillate approximately in the range of the red double-arrow.}
\label{fig_sketch1}
\end{figure}

We can gain some insight into the possibility of oscillatory behavior in a given system
by considering its flow curve, namely the relation between the stress and the velocity of deformation, under stationary conditions.
In standard depinning or yielding models this curve has a typical form as depicted in Fig. \ref{fig_sketch1}(a). There is a univocal relation between driving velocity $V_0$ and stress $\sigma$, that starts at a finite stress value $\sigma_c$ when $V_0\to 0$. In this situation, if the system is driven at constant and very small $V_0$, the stress remains constant (up to statistical fluctuations) and very close to the critical value $\sigma_c$. In these conditions the avalanche size distribution shows a cutoff power law, with no king events.
In the case in which the flow curve has a reentrance as shown in \ref{fig_sketch1}(b) the behavior of the system becomes more subtle. Actually, in this case the possibilities of driving at constant stress, or constant velocity must be carefully distinguished. If driving the system at constant stress in the range $\sigma_0$-$\sigma_c$, the system may be either at rest, or evolving at a finite velocity. If driven at constant small velocity instead, and assuming a stationary situation, the stress should be very close to $\sigma_c$.
However, when the flow curve has negative slope ($dV_0/d\sigma<0$), a mechanical stability analysis indicates that a stationary solution may become unstable, and the system may enter a more complicated non-stationary temporal dynamics, in which large king avalanches can appear, and the stress oscillates in the range $\sim\sigma_0$-$\sigma_c$. The condition $dV_0/d\sigma<0$
on a flow curve is termed a {\em velocity weakening} condition in the seismic literature\cite{dieterich,ruina}, and is considered as one of the most important characteristics of systems that are able to realistically describe earthquakes\cite{scholz}.

While a qualitative understanding of the origin of king avalanches and the possibility of oscillatory global behavior in system with velocity weakening characteristics  has been achieved, detailed behavior in particular models have been more difficult to obtain. 
For instance, the way in which a system that displays a smooth stationary dynamics (i.e., normal depinning behavior) for some values of its defining parameters, transitions to a synchronized system with quasi-periodic king avalanches for other values of the parameters is rather unclear\cite{petrillo2025anomalous,landes,wyart1,wyart2}. Our presentation addresses precisely this kind of question. We present a simple model that is built upon the  
paradigm of depinning systems, with the addition of an ingredient associated to ``aging" that we show produces a reentrant behavior of the flow curve. This in turn leads to the possibility of global instabilities that give rise to the appearance of king avalanches. We construct the mean field phase diagram of the system as a function of two main controlling parameters: the stiffness of a driving spring $k_0$ and the driving velocity $V_0$. We investigate the different dynamical regimes in this parameter space, and the transitions among them. After that, we analyze the situation in the case of nearest neighbor interactions in two-dimensional systems.

\section{Model}

The model we study is based on a particular implementation of the depinning problem\cite{fisher1998collective,kardar1998nonequilibrium,wiese2022theory}. We consider the evolution of a set of variables $u_i$. Here $i$ defines the spatial position, so for instance in a two-dimensional geometry it must be understood that $i$ labels both $x$, and $y$ coordinates: $i\equiv (x,y)$. The coordinate $u_i$ evolves in an overdamped way under the action of all the  forces acting on it:
\begin{equation}
\lambda \frac{du_i}{dt}=(w-u_i)k_0 +\sum_j G_{ij}(u_j-u_i)-\frac{dV_i(u_i)}{du_i}.
\label{fundamental}
\end{equation}
The first term of the r.h.s. is the driving force. It is represented by a spring of stiffness $k_0$ joining the position $u_i$ to a driving point $w$ which is common to the whole system. In general  we will use $w=V_0 t$, which defines the driving velocity $V_0$. The average value of this term allows to define the average stress on the system $\sigma$:
\begin{equation}
\sigma\equiv N^{-1}\sum_ i (w-u_i)k_0
\end{equation}
where $N$ is the number of variables we have in the system.

The middle term in  (\ref{fundamental}) represents the coupling between different sites. It is written in a way in which $G_{ij}$ can be interpreted as a spring constant connecting sites $i$ and $j$. We will study two particular cases in detail. One is the {\em mean field} case,
in which $G_{ij}$ is the same for all pairs $( i,j)$:  $G^{MF}_{ij}\equiv k_1/N$, with an overall intensity given by the stiffness $k_1$. The second case corresponds to a nearest neighbor interaction in a two dimensional square lattice with periodic boundary conditions. In this case $G^{NN}_{ij}\equiv k_1/4$ if $i$ and $j$ are nearest neighbors, otherwise $G^{NN}_{ij}=0$. These two cases will serve as prototypical examples of a fully interconnected lattice (MF), and a case with only short range interactions (NN).

\begin{figure}
\includegraphics[width=8cm,clip=true]{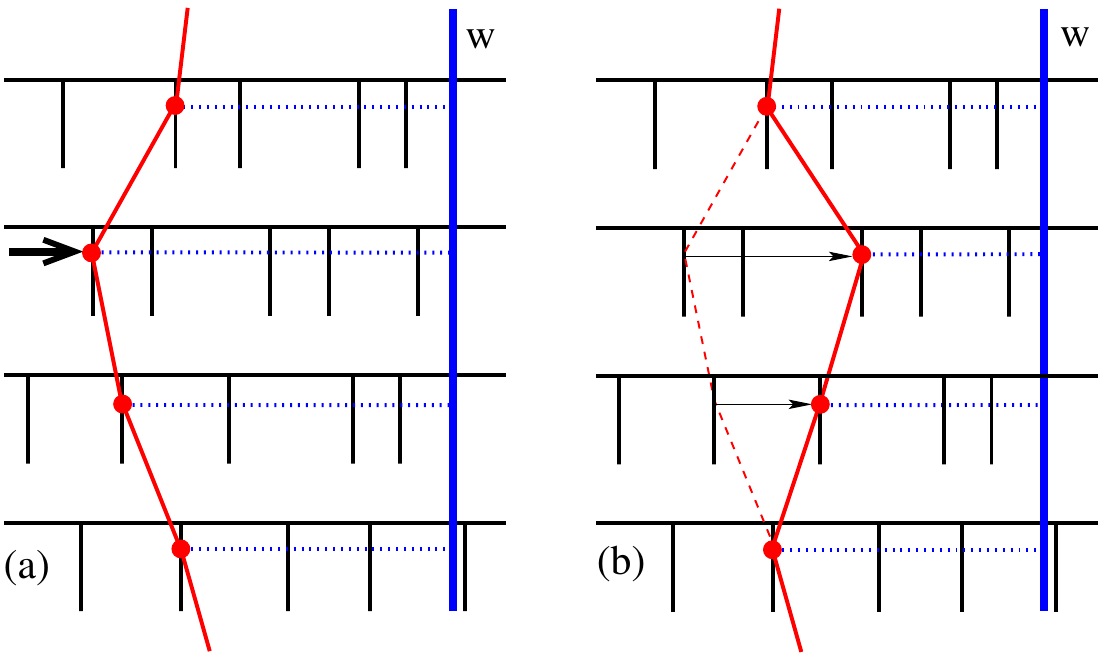}
\caption{Scheme of an avalanche taking place in a one-dimensional version of the model. An initially unstable site (indicated by the arrow in the left panel) triggers an avalanche that--in this example--destabilizes two sites, and leads to the final configuration in panel (b). Note that $w$ is fixed during the process.}
\label{fig1}
\end{figure}

Finally, the last term in (\ref{fundamental}) is a pinning force generated by some disorder potential $V_i$ acting on the coordinate $u_i$. Many previous investigations have shown that the exact form of the potentials $V_i$ are not crucial in determining the global properties of the system, as long as a few general characteristics are maintained. In particular, the form of the potentials must be uncorrelated between different sites, and on a given site $V_i(u_i)$ has to be uncorrelated between different, well separated values of the argument. Within these restrictions, we choose a form for the potential $V_i$ referred to as the narrow wells approximation\cite{abbm1,abbm2,jagla2014viscoelastic}. It corresponds to have a collection of infinitely narrow wells, randomly distributed along the coordinate $u_i$ with an average separation $d$, in such a way that if the variable $u_i$ is within one of these wells, it takes a force $f_0$ to take it out from it.
In addition, we will assume that if a coordinate $u_i$ goes out of a given well, it immediately reaches the next well available, where it sits until the applied force is sufficient to make it escape from it. These rules transform the continuous time dynamics of Eq. (\ref{fundamental}) in a discrete evolution that takes the form of a cellular automaton. 
Concretely, for a given value of time $t$, we calculate the force $f_i$ acting on the variable $u_i$ as the two first terms on the r.h.s. of Eq.  (\ref{fundamental}):

\begin{equation}
f_i\equiv (w-u_i)k_0 +\sum_j G_{ij}(u_j-u_i).
\label{f}
\end{equation}
If $f_i>f_0$ for some $i$'s, the corresponding $u_i$'s are moved to the next well according to
\begin{equation}
u_i\to u_i + \Delta u_i
\end{equation}
where $\Delta u_i$ are taken from an exponential distribution of mean $d$ (the exponential form is a consequence of the random placement of wells along the $u$ coordinate). These changes in $u_i$ may produce a further instability of other sites (detected through a re-evaluation of Eq. (\ref{f})), and a cascade effect producing an {\em avalanche} (see Fig. \ref{fig1}). When no unstable sites remain, we compute the size of the avalanche $S$ as
\begin{equation}
S=\sum_i{\Delta u_i}
\label{S}
\end{equation}
where $\Delta u_i$ is the total displacement of $u_i$ during the avalanche. Note that according to the definition of the stress $\sigma$, an avalanche of size $S$ corresponds to a stress drop $\Delta\sigma$ in the system of value
\begin{equation}
\Delta\sigma=k_0S/N.
\label{delta_sigma0}
\end{equation}

Once the avalanche has stopped, the driving $w$ is increased until the force onto some site reaches $f_0$, and a new avalanche is triggered. We emphasize that $w$ is kept fixed during the development of an avalanche. This is equivalent to say that avalanches are  {\em instantaneous} in the time scale of driving, and in this sense, up to here the evolution of the system is totally independent of the value $V_0$. In this situation, the flow curve of the system, i.e., the $V_0-\sigma$ dependence degenerates into a vertical line defining the position of the critical stress $\sigma_c$.

The model just presented provides a possible implementation of the depinning transition. The characteristics of the avalanches that this model displays 
can be summarized as follows. The avalanche distribution has a cutoff power law form, namely
\begin{equation}
P(S)\sim S^{-\tau_0}\exp(-S/S_{max})
\end{equation}
The value of $\tau_0$ is $\tau_0=3/2$ for MF interactions, and $\tau_0\simeq 1.27$ for two-dimensional NN interactions. The value of $S_{max}$ increases as $k_0$ decreases, in the form
\begin{equation}
S_{max}\sim k_0^{-\sigma}
\end{equation}
with $\sigma=2$ for MF, and $\sigma \simeq 1.38$ for two-dimensional NN interactions.
As we already indicated, while taking the avalanche duration as negligibly small, the behavior of the system is independent of the value of $V_0$. The system does not display any sign of king avalanches, or global synchronization.

Now we introduce a modification in the model intended to mimic 
 well known effects that are documented in the literature. For instance\cite{carpick}, for some materials of interest in relation to seismic faults, an AFM tip tends to become more bounded to a substrate as a function of the time-into-contact, requiring a larger forces to displace it for larger contact times.
We model this effect by making the value of the depinning force of a given well to depend on the time that the particle spent in that well, that we note $T$. The depinning force is set to be some value $f_0-\beta$ initially (at $T=0$), evolving asymptotically to the stationary value $f_0$ as $T\to\infty$.
For concreteness we take this evolution to have an exponential dependence on $T$, and note it as $F_0(T)$:
\begin{equation}
F_0(T)= f_0-\beta \exp(-T/\tau).
\label{f0det}
\end{equation}
In this way, $\beta$ measures the intensity of the aging effect, and $\tau$ is a fixed time scale that will control the velocity of the aging. Notice that now the behavior of the system will depend on the value of the aging time $\tau$ compared to the time scale set by the driving velocity $d/V_0$. Therefore, now the properties of the system will depend on the value of the parameter $V_0\tau/d$. 
In Appendix I we show through a simplified analysis how this kind of effect leads in a one degree of freedom system to a dynamics that can be smooth or stick-slip depending on the relative values of $k_0$, $\beta$, and $V_0\tau/d$. In concrete, if $V_0>\beta d/(k_0\tau)$, the body moves smoothly at the driving velocity $V_0$, with the $k_0$ spring pulling with a force $\sigma=f_0-\beta$. However, if $V_0<\beta d/(k_0\tau)$, there is no stationary solution, and the body enters a stick slip behavior, alternating between rest periods (in which $\sigma$ increases linearly in time) and abrupt displacements of the body with associated decreases of $\sigma$. The border separating the two regimes in this simplified analysis is $V_0=\beta d/(k_0\tau)$, which will serve as a reference when we construct the phase diagram of the extended system.

\section{Results for Mean Field Interactions}

When interactions are taken to be mean field, 
some exact analytic results can be obtained that complement the results of numerical simulations. We separate the results by first analyzing the pure depinning case ($\beta =0$) and then presenting the case with relaxation ($\beta \ne 0$).

\subsection{Results without relaxation ($\beta= 0$)}

This corresponds to the standard depinning case without aging. This case has been very well studied, and we simply re-derive here some of the main results to serve as reference for the case in which aging is included.
For a system of very large size, the state of the system at a given time is completely defined once we know the distribution function of different values of the positions $u$. We note this distribution as ${\cal P}(u)$.
We assume from the beginning that under a uniform driving ($w=V_0t$) the dynamics of the model is stationary, namely that up to stochastic fluctuations (due to the stochastic positions of the pinning wells) the average coordinate $\overline u$ proceeds also at the same constant velocity $V_0$.
 This means that the distribution ${\cal P}(u)$ will have a fixed form that simply shifts to increasing values of $u$ with velocity $V_0$ as $t$ increases.
The form of ${\cal P}(u)$ can be determined at once. First notice that 
there is a particular value of $u$ at which the force will be exactly $f_0$, and therefore
${\cal P}(u)=0$ if $u<u_0$ as no site can be stable under an applied force larger than $f_0$. The value of $u_0$ is determined
by the condition
\begin{equation}
f_0=k_0 (w-u_0)+ (\overline u -u_0)k_1
\label{f0}
\end{equation}
For $u> u_0$ the form of ${\cal P}(u)$ is simply obtained observing that a site could have reached a particular value of $u$ if in its last jump it went from some potential well at the left of $u_0$ to the next potential well at $u$. As wells are randomly placed with average distance $d$, this gives an exponential distribution which is the form that ${\cal P}(u)$ will follow. Namely
\begin{eqnarray}
{\cal P}(u)&=&{d}^{-1}\exp(-(u-u_0)/d) ~~~ \mbox {if} ~~~u>u_0\\
{\cal P}(u)&=&0 ~~~~~~~~~~~~~~~~~~~~~~~~~~~~~~\mbox {if} ~~~u<u_0 
\end{eqnarray}
The value of $u_0$ is calculated by evaluating $\overline u$ from this expression, and inserting into Eq. (\ref{f0}). It results
\begin{equation}
u_0=w-(f_0-k_1d)/k_0.
\label{u0}
\end{equation}
In addition, the value of the stress $\sigma=k_0(w-\overline u)$ can also be easily calculated and results
\begin{equation}
\sigma=f_0-(k_0+k_1)d.
\end{equation}

The calculation of the avalanche size distribution requires considering the actual distribution of values $u_i$ within the average distribution ${\cal P}(u)$. The main point is that because of the uncorrelated nature of the potentials, the actual values of $u_i$ are randomly placed,  and the development of an avalanche can be mapped to the problem of the return to zero  of a random walk with a bias proportional to $k_0$ (see Appendix II).
A nice closed form for the size distribution of avalanches can be obtained when they involve many sites ($S\gg d$). The result is
\begin{equation}
P(S)\sim S^{-3/2}e^{-S/S_{max}}
\end{equation}
with ${S_{max}=2d(k_0+k_1)k_1/{k_0^2}}$
In addition, it has to be mentioned that the assumed temporal homogeneity of the dynamics is always fulfilled in the absence of aging, and that avalanche size distribution follows this formula for any non-vanishing value of $k_0$.

\subsection{Results with relaxation ($\beta> 0$)}

Now we consider the model in the presence of aging ($\beta>0$). It turns out that this case is far richer than the $\beta=0$ case. We will start the analysis assuming that the dynamics is stationary in time (but we anticipate that this will not always be the case). The first main goal is to calculate the value of $\sigma$ in the system, which in the present case will depend on the value of the driving velocity $V_0$. The analysis is made along the same lines of the $\beta=0$ case, however, now it is not enough to consider the global distribution ${\cal P}(u)$ as we did previously. In fact, to determine when a variable $u_i$ becomes unstable, we  need information on how much time it spent in that particular well, since this determines the (now time dependent) pinning force $F_0(T)$ (Eq. (\ref{f0det})). It will therefore be convenient to consider a two variable distribution 
${\cal P}(u,T)$, in such a way that ${\cal P}(u,T)dudT$ is the number of sites that have positions in $(u,u+du)$, and have been in their corresponding well during a time between $T$ and $T+dT$. 

\begin{figure}
\includegraphics[width=6cm,clip=true]{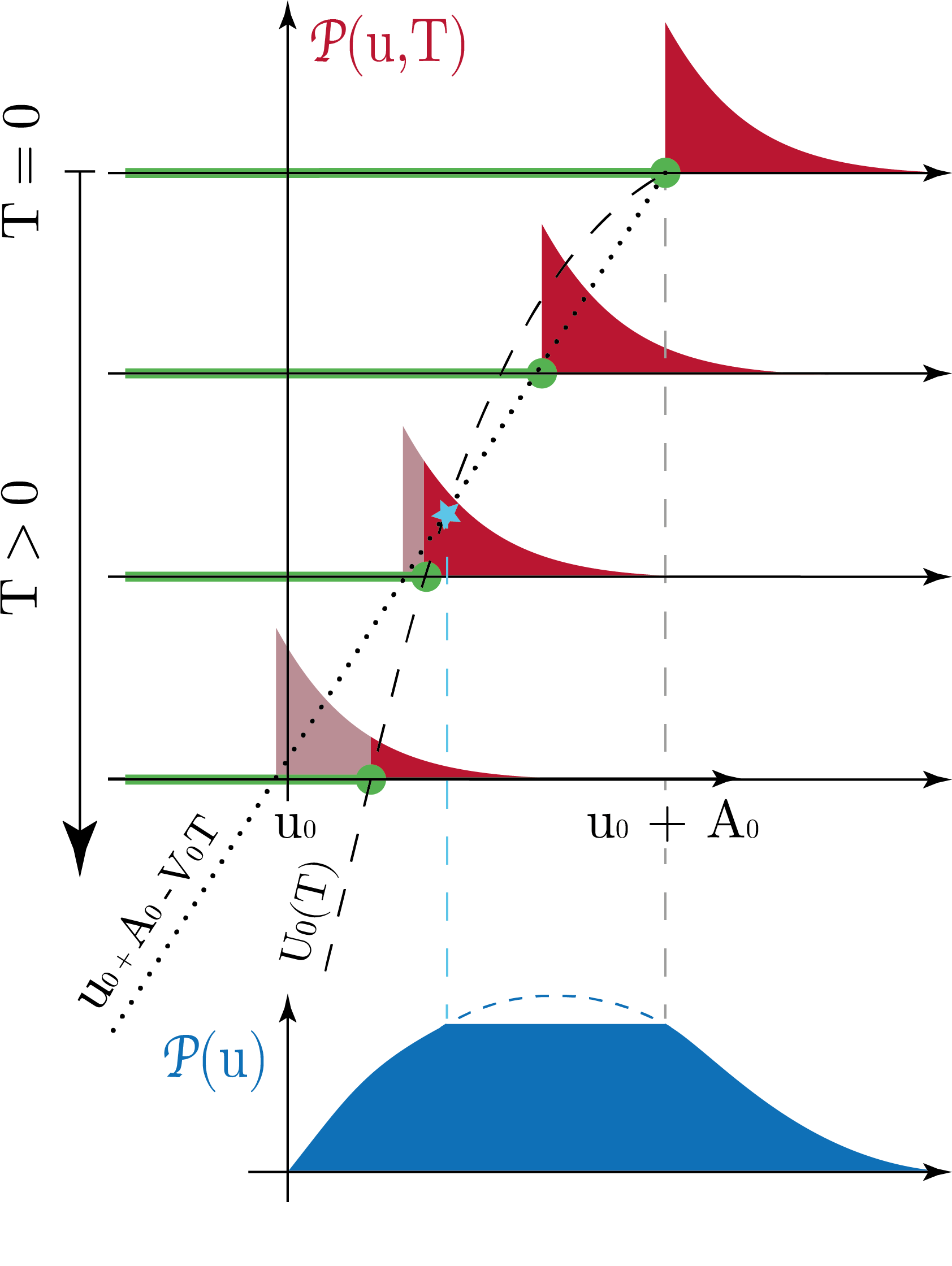}
\caption{The form of ${\cal P}(u,T)$ in the presence of aging ($\beta> 0$). The minimum value $U_0(T)$ of stable positions for each $T$ is indicated by the dashed line. In the last panel, the complete ${\cal P}(u)=\int dT {\cal P}(u,T)$ is shown. Notice that a situation with $V_0\tau<A_0$ has been assumed.
}
\label{put}
\end{figure}

\begin{figure}
\includegraphics[width=8cm,clip=true]{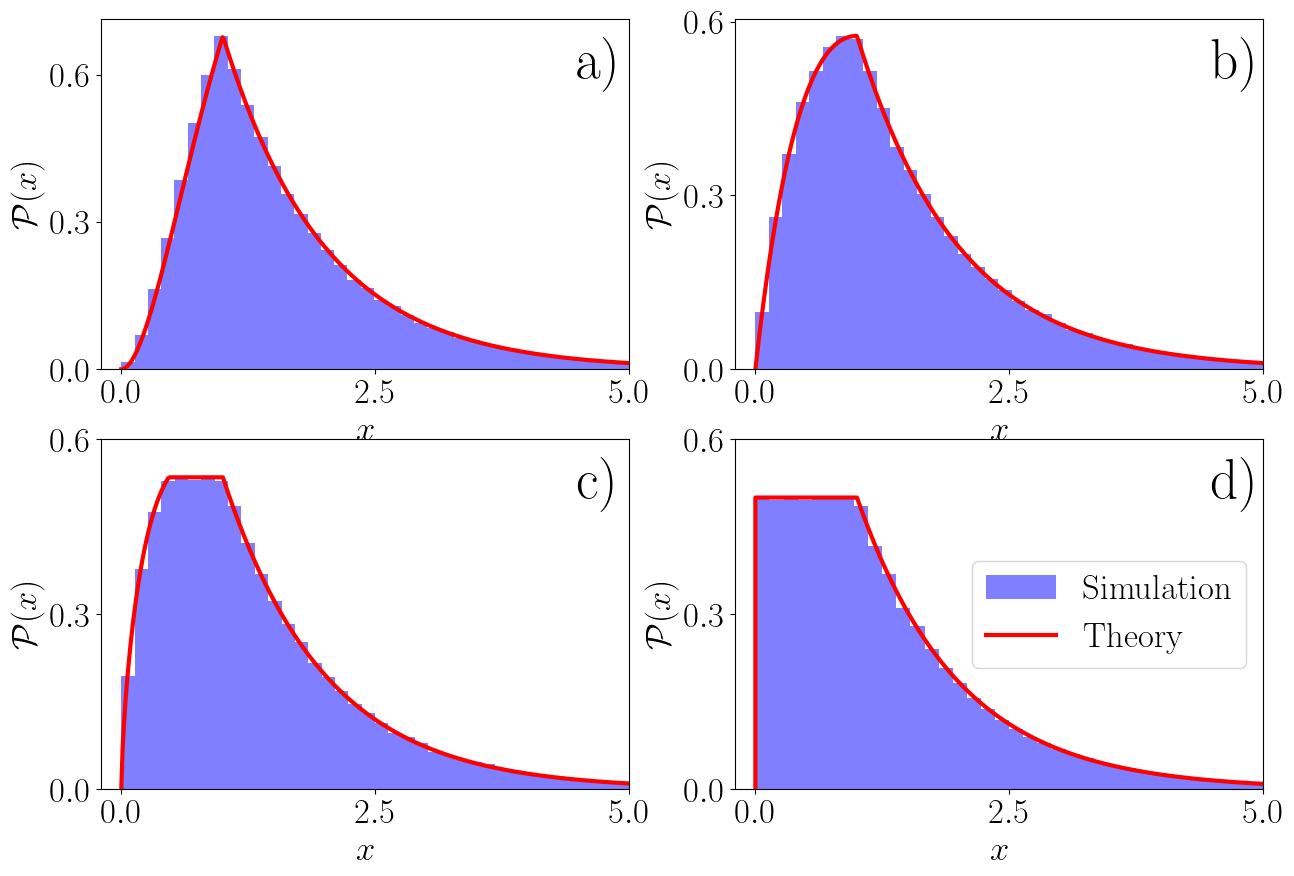}
\caption{Comparison of analytical an numerically obtained forms of ${\cal P}(u)$, in the case the dynamics is smooth. Parameters are $\beta=0.2$, $k_1=0.1$, $k_0=0.1$. Values of $V_0\tau/d$ are 2, 1, 0.7, and 0, from a) to d).}
\label{fig:dist_general}
\end{figure}

\begin{figure}
\includegraphics[width=8cm,clip=true]{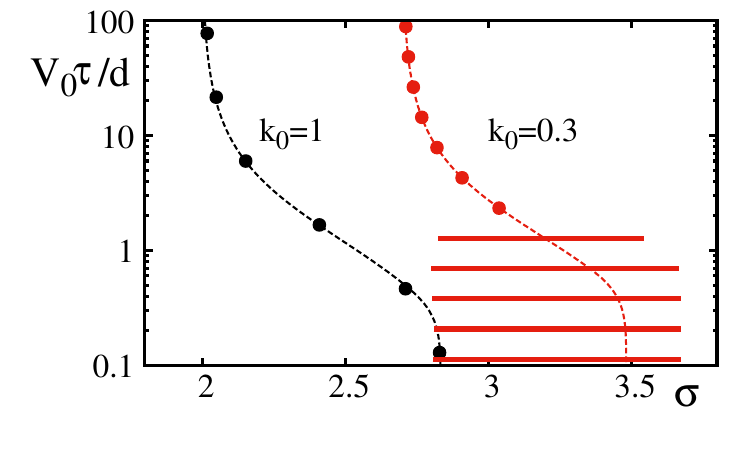}
\caption{Dotted lines: analytical $V_0$-$\sigma$ dependence in the system for two different  values of $k_0$. Other parameters are $k_1=1$, $\beta=1$. Note that the analytical result is obtained assuming that the dynamics is stationary in time. Points and segments are the result of numerical simulation in a system of $N=256^2$ sites. Segments indicate the range in which stress oscillates along time, when the dynamics is non-stationary.} 
\label{comparison}
\end{figure}

To obtain the form of ${\cal P}(u,T)$, let us refer to Fig. \ref{put}.
In the case $\beta=0$, there was a minimum value $u_0$ of $u$, such that particles are at stable position only if $u>u_0$. In the present case, the minimum value of $u$ depends on $T$, as the maximum pinning force of a well does. We call $U_0(T)$ this minimum value, which generalizing expression (\ref{f0}) can be obtained from the condition
\begin{equation}
F_0(T)=(w-U_0)k_0+(\overline u -U_0)k_1
\end{equation}
and using (\ref{f0det})
\begin{equation}
U_0(T)=u_0+A_0\exp{(-T/\tau)}
\end{equation}
with $u_0$ given by Eq. (\ref{u0}), and $A_0={\beta}/(k_0+k_1)$.
The form of $U_0(T)$ is plotted in Fig. \ref{put} with a dashed line. Consider the particles that have just destabilized ($T=0$) and find new positions along the $u$ axis. They can only stay at wells which are located at $u>U_0(0)=u_0+A_0$.  For $u$ values larger than this, the distribution decays exponentially because the wells are randomly located. 
If from this configuration a small time $T$ is elapsed, ${\cal P}(u,0)$ must transform into ${\cal P}(u-V_0 T,T)$ (this shift of ${\cal P}(u,T)$ with $T$ is indicated in Fig. \ref{put} by the dotted line). However, in this process, it may be necessary to trim the distribution to avoid sites being at the left of $U_0(T)$. This is graphically indicated in Fig. \ref{put}, with trimmed region shown in lighter color. This analysis leads to the conclusion that the form of ${\cal P}(u,T)$ can be written (up the a normalizing factor) as
\begin{widetext}
\begin{equation}
    {\cal P}(u,T)\sim e^{\displaystyle -(u-u_0-A_0+V_0 T)}  \Theta \left(\max \left[u-u_0-A_0+V_0 T  ,  u-U_0(T)\right]) \right)
    \label{pdeut}
\end{equation}
\end{widetext}
($\Theta$ is the Heaviside step function).
From this expression, ${\cal P}(u)=\int_0^{\infty} {\cal P}(u,T)dT$ is obtained through elementary integration, yielding the result  ($x\equiv u-u_0$)
\begin{equation}
{\cal P}(u) \propto
\begin{cases}
0, & \text{if } x < 0, \\[6pt]
\min\left [ 1, \left( \dfrac{x}{A_0} \right)^{V_0 \, \tau} e^{-(x - A_0)}    \right ], & \text{if }0 \leq x < A_0, \\[6pt]
e^{-(x - A_0)}, & \text{if } x \geq A_0,
\end{cases}
\label{eq:dist_general}
\end{equation}
This distribution is plotted in the last panel of Fig. \ref{put}, and in Fig. \ref{fig:dist_general} for different values of $V_0 \tau/A_0$.
Note that  an intermediate region with ${\cal P}(u)$ constant appears only in the case where $V_0 \tau<A_0$.
From ${\cal P}(u)$, the mean value $\overline u=\int_0^\infty {\cal P}(u)du$ can be computed, and replaced into (\ref{f0}) to obtain the value of $\sigma$. The dependence of $\sigma$ on $V_0$ obtained in this way is plotted in Fig. \ref{comparison}.
The most remarkable feature of this dependence is its velocity weakening characteristic, i.e, the value of $\sigma$ decreases as $V_0$ increases, as it was qualitatively anticipated to occur.

It must be stressed that these results were obtained on the {\em assumption} that the dynamics is stationary in time. The finding of a velocity weakening behavior is however a warning that instabilities may occur, leading to some king of oscillatory behavior of the system. 

\begin{figure}
\includegraphics[width=8cm,clip=true]{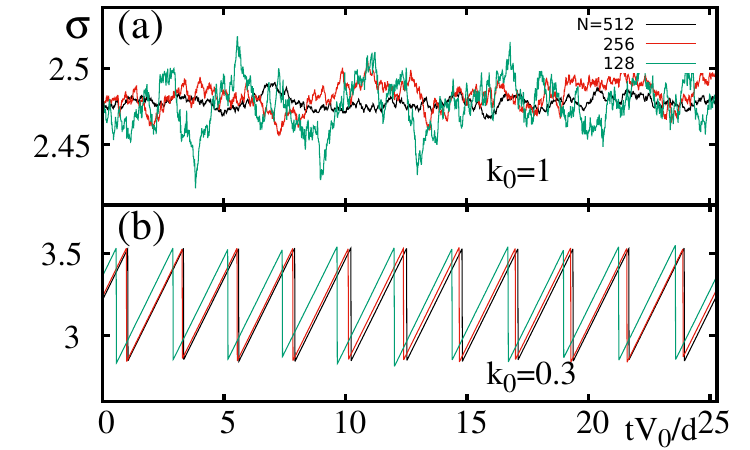}
\caption{(a) The form of $\sigma(t)$ at a large (a) and small (b) value of $k_0$ in system of different sizes. Other parameters are: $k_1=1$, $\beta=1$, $V_0\tau/d=1.274$. 
In (a), the fluctuations decrease with system size. In (b) there is no dependence on the system size, and the dynamics is clearly non-stationary.}
\label{detalle}
\end{figure}

We will investigate first the possibility of oscillatory behavior by resorting to direct numerical simulation.
They were performed in a system of $N=256^2$ sites, that evolves according to the rules explained in Section II.
The control parameters are the values of $k_0$, $k_1$, and $V_0$. The value of $\beta$ will be fixed at $\beta=1$. 
There are parameters for which we observe a smooth dynamics (Fig. \ref{detalle}(a)), and in this case the result obtained for $\sigma$ coincides with the previously obtained analytic one (points in Fig. \ref{comparison}). In this case, the numerically obtained ${\cal P}(u)$ matches nicely 
the one obtained analytically (Fig. \ref{fig:dist_general}).
However for other values of the parameters (in particular, if $k_0$ is sufficiently reduced), the time evolution of $\sigma$ is clearly non-stationary (Fig. \ref{detalle}(b), and segments in Fig. \ref{comparison}). Remarkably, we find also parameter values for which the dynamics can be either stationary, or non-stationary, depending on details of the initial conditions.

\begin{figure}
\includegraphics[width=9cm,clip=true]{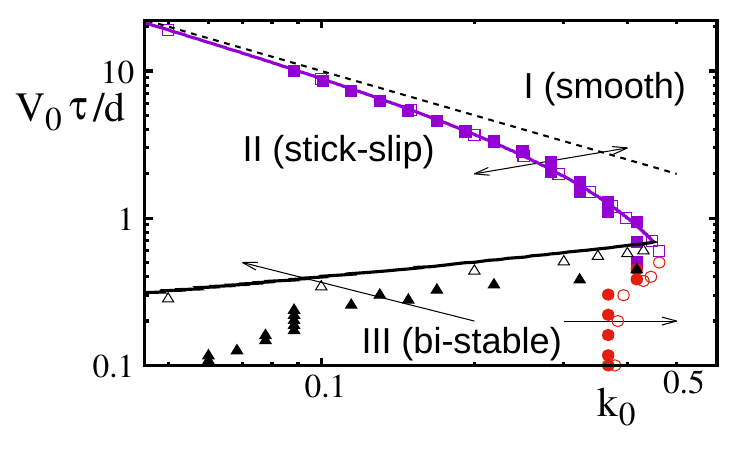}
\caption{Phase diagram of the system in the  $k_0$-$V_0\tau/d$ plane (we use $k_1=1$, $\beta=1$). 
Full and open symbols indicate the transition between different regions determined by two different schemes of numerical simulations as explained in the text. The continuous line is the value of $k_{0c}$ at which $S_{max}$ diverges (from a linear stability analysis of the smooth phase), then defining the 
border of the II region, where only a non-smooth dynamics is possible. 
Dashed line is the separation between smooth and stick-slip phases of the one-variable model of Appendix I.
}
\label{phase_diagram}
\end{figure}

By performing systematic numerical simulations for different set of parameters, we were able to obtain a phase diagram of the system, that we present as a function of the parameters $k_0$ and $V_0\tau/d$ in Fig. \ref{phase_diagram}. The results of the simulations are represented by the points, which separate different qualitative behaviors of the system, that are describes below. Before that, we mention that the full symbols correspond to straightforward simulations (as explained above) in systems with $N=256^2$, while open symbols are the results or more elaborate simulations in which the exponential dependence of ${\cal P}(u,T)$ (Fig. \ref{put}) is analytically taken into account. This means that formally, open symbols correspond to results for systems in the thermodynamic limit ($N\to\infty$). All in all, up to numerical uncertainties empty and full symbols are consistent with each other and delineate the three regions of the phase diagram that we now describe.

\begin{figure}
\includegraphics[width=8cm,clip=true]{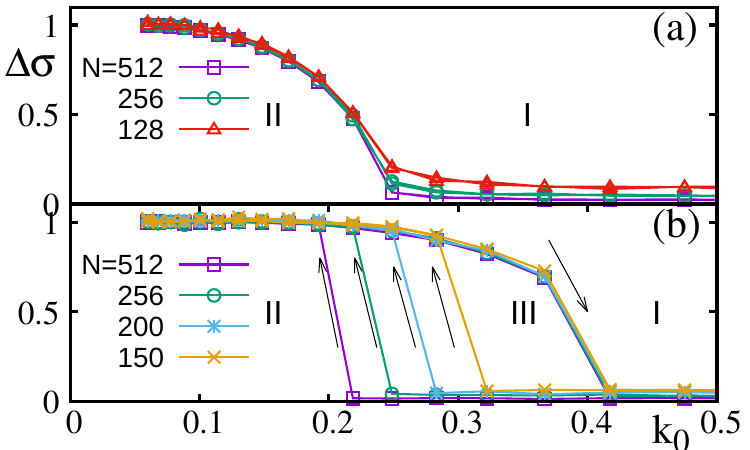}
\caption{Standard deviation of the $\sigma(t)$ signal as a function of $k_0$, displaying the II-I reversible transition (panel (a), $V_0\tau/d=3$ and the hysteretic II-III-I transition (panel (b), $V_0\tau/d=0.3$. Curves for different system sizes show the convergence to a non-fluctuating value in the thermodynamic limit of the phase I.
}
\label{delta_sigma}
\end{figure}

Fig. \ref{phase_diagram} displays three regions of  different
qualitative behavior. At large values of $V_0\tau/d$ and $k_0$ we find a region (labelled I) in which the only observed dynamical state of the system is a stationary state with constant $\sigma$. However, if the values of $k_0$ and/or $V_0\tau/d$ are sufficiently reduced, the simulation shows that the smooth dynamics becomes unstable, and the system enters a stick-slip regime. This is an expectable result in view of the qualitative analysis presented in Appendix I, and it is also a result that has been observed in similar models. However, we find here an unexpected interesting situation. For some parameters (region II), the stick slip behavior is the only dynamical state observed, no matter the initial conditions or the history of parameter evolution that led to the actual set of parameters. For other parameter values (those in region III)
we may have  either a smooth dynamics, or a stick-slip one, depending on how the system is initially prepared, or how is the history of parameter changes that led to the actual values of $k_0$ and $V_0\tau/d$. For instance, if we change smoothly the parameter $k_0$ and cross the II-I border 
we observe a smooth transition between stationary and non-stationary dynamics (see Fig. \ref{delta_sigma}(a)). In this case the amplitude of the oscillation in $\sigma$ starts being very small at the transition point and increases away from it. This is reminiscent of a Hopf bifurcation in dynamical systems \cite{strogatz}. In the case we choose a value of $k_0$ following a II-III-I path
(Fig. \ref{delta_sigma}(b)), the transition is seen to be hysteretic, and the possibility to have either a stationary or non-stationary dynamics in the intermediate region is clearly observed. This behavior can be interpreted in the language of dynamical systems as a saddle-node bifurcation of cycles\cite{strogatz}.

Although the existence of oscillatory dynamical states is a robust outcome of the numerical simulations, it would be worth having some analytical description of it as well.
The analytical calculation presented previously has assumed that the dynamics is stationary in time, and it cannot be simply extended to consider the possibility of stick-slip behavior. However, we can do some sort of stability analysis onto the smooth solution to obtain the stability limits of this region and therefore the borders  of the region of the phase diagram where a smooth dynamics is possible. 
The analysis is based on the consideration of the avalanche size distribution $P(S)$, and is fully developed in Appendix II. The conclusions are as follows.

Within the case of mean field interactions, at any values of the parameters corresponding to a smooth behavior of the system, the avalanche size distribution has the form
\begin{equation}
P(S)\sim S^{-3/2}\exp(-S/S_{max})
\end{equation}
with $S_{max}$ depending on the parameters of the system. The fact that a cutoff at 
$\sim S_{max}$  (independent of $N$) exists in the size distribution is a confirmation that the dynamics is smooth. In fact, it tells that the maximum stress drop an avalanche can produce (Eq. \ref{delta_sigma0}) becomes negligible as system size increases, then justifying the constant-in-time value of $\sigma$ in the thermodynamic limit. 
If for particular parameter values we have $S_{max}\to\infty$ this would indicate a potential instability of the smooth dynamics, as an infinitely large avalanche (more properly, an avalanche comparable to the size of the system) will produce a finite change in $\sigma$, and therefore the assumption of a constant-in-time $\sigma$ value does not hold. 
The stability analysis then resorts to the calculation of $S_{max}$.
In the absence of relaxation $S_{max}$ turns out to be (see Appendix II) ${S_{max}=2d(k_0+k_1)k_1/{k_0^2}}$, and this is finite for any non-zero value of $k_0$, justifying the conclusion that the smooth dynamics is always stable in the absence of relaxation. The analysis of the case with relaxation shows instead that while $S_{max}$ is finite for sufficiently large $k_0$, it increases and diverges at some finite critical value of $k_{0c}$, which therefore represents the theoretical stability limit of the smooth phase. 
The value of $k_{0c}$ as a function of $V_0\tau/d$ (which is obtained in the Appendix II) is plotted in Fig. \ref{phase_diagram} to compare with the numerical results. We see in fact that it reproduces the borders of the phase diagram in which the smooth phase is no longer stable, namely the reversible I $\leftrightarrow$ II transition, and the irreversible III $\to$ II transition.

\section{Results for nearest-neighbor Interactions in two-dimensions}

The all-equal strength elastic interaction studied in the previous section is an idealized situation in which analytical results can be obtained, and therefore serves as a reference and a starting point for the study of more realistic interactions. When going to systems with distance dependent interactions, spatial dimensionality becomes relevant. In the present study we limit our analysis to two-dimensional system which are, on one side, realistic from the point of view of friction, seismic phenomena, etc, and on the other side avoid typical marginal behavior that may occur in the study of one dimensional systems. With respect to the dependence of the elastic interaction with distance, we concentrate here in the simplest nearest neighbor interaction over a square lattice, which simplifies a great deal the numerical simulation. However we emphasize that elastic interactions decaying with distance with some power are very relevant in applications, but also more demanding numerically, and we left those cases aside for the moment. We maintain potentials in the narrow well approximation, in such a way that a simulation in terms of ``cellular automata" still applies. The description we did in Section II of the numerical scheme continues to be valid, with the only difference that we take for the elastic kernel $G_{ij}$ (Eq. (\ref{f})) the appropriate form for nearest neighbor interaction $G_{ij}^{NN}$. This transforms Eq. (\ref{f}) into
\begin{equation}
f_i\equiv (w-u_i)k_0 +\frac {k_1} 4{\sum_{j} }' (u_j-u_i).
\end{equation}
where the prime in the sum indicates that this is restricted to the four neighbors of site $i$.

The synchronization observed in the case of MF interactions for some parameter range is expected to weaken in systems with NN interactions. This is naturally associated to the fact that the mutual influence between two far away sites proceeds now through chains of sites in between them, contrary to the MF case where there is a direct interaction between any pair of sites. This rises the fundamental question if global synchronization may exist at all in systems with short range interactions. 
There are well known cases of closely related models\cite{landes} in which the MF synchronization completely disappears in the case of short range interactions. 
Actually, although the possibility to have global syncronization of a system of friction oscillators with local interactions has been proposed and qualitatively studied by Sornette {\em et al.} \cite{Smithetal1994,Osorio2010}, concrete models based on microscopic dynamical rules displaying this behavior have not been considered in detail.

The central question we want to address in the case of the present model with NN interaction is: are there parameters for which the system globally syncronizes, even in the limit $N\to\infty$?
We will see that numerical simulations strongly points to the fact that this syncronization is indeed possible. This is actually one of the main results of our work.
However, we will find that, for the system under analysis, this synchronization does not involve the existence of system size scaling (king) avalanches.

We investigated in detail the behavior of a system  at a fixed value of $V_0\tau/d=10$ \textcolor{red}, when $k_0$ is varied. The analysis was performed for systems of different sizes $N=L\times L$, to investigate the stability of the dynamical states encountered as $L$ increases. For the present NN case, we took a value of $k_1=0.1$, which displays more clearly the expected behavior. The observed behavior is robust against changes of this parameter.

\begin{figure}
\includegraphics[width=9cm,clip=true]{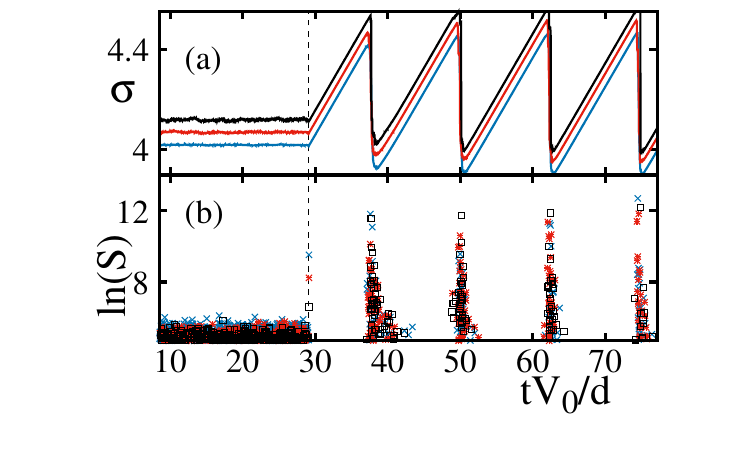}
\caption{(a) The form of $\sigma(t)$, at $k_0=0.1$ (at the left) and $k_0=0.05$ (at the right of the vertical dashed line), with $V_0\tau/d=8.541$ and different system sizes $N=L\times\L$, with $L=128$ (upper curve, squares), $L=256$ (medium, stars),  and $L=512$ (lower, crosses) (in (a) curves have been shifted vertically, for clarity).  (b) Corresponding avalanches in the system as a function of time. Note that in order to have a similar temporal density of avalanches only those occurring in an area of $128\times 128$ are reported in the three cases.}
\label{sigma_sr}
\end{figure}

\begin{figure}
\includegraphics[width=9cm,clip=true]{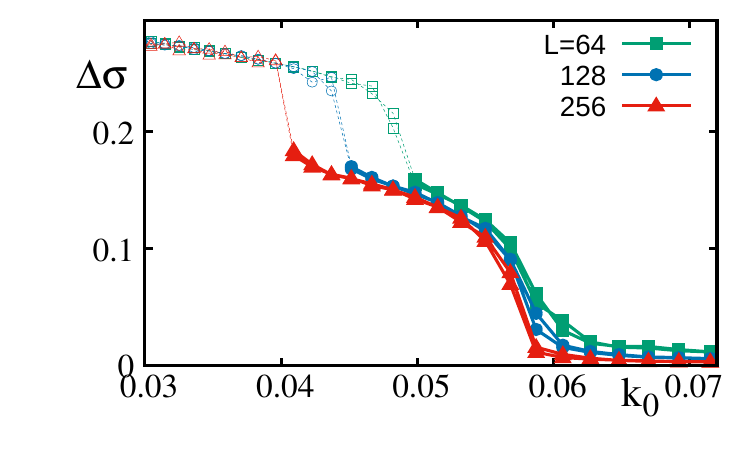}
\caption{Standard deviation of $\sigma(t)$ in simulations increasing and decreasing $k_0$ for different system sizes $N=L\times L$ at $V_0\tau/d=10$, $k_1=0.1$. Data point to a sharp transition at $k_0\simeq 0.058$ in the limit $L\to\infty$. The part of the curves
with open symbols and dashed lines displays a transition to a fully synchronized state at very low $k_0$ that is deemed to be a system size effect.}
\label{transition_sr}
\end{figure}

\begin{figure}
\includegraphics[width=8cm,clip=true]{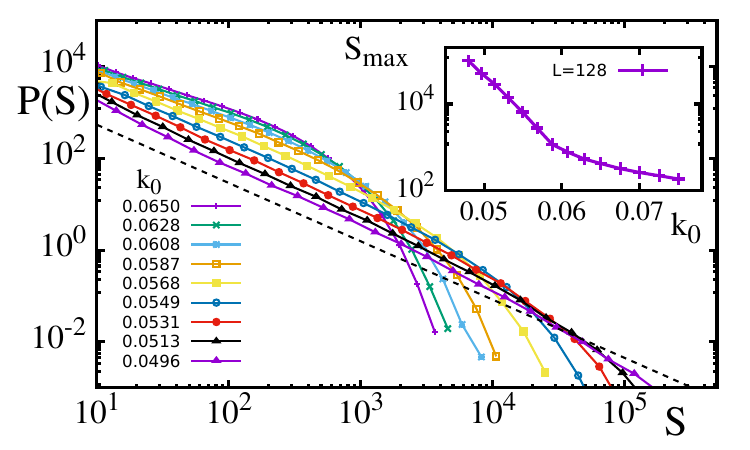}
\caption{Avalanche size distributions as a function of $k_0$ for a system with $L=128$. For reference, dashed line has a slope -1.27. Inset: $S_{max}$ vs $k_0$ for the curves in the main plot. Results in this figure do not change by increasing size, and are representative of the thermodynamic limit.}
\label{size_sr}
\end{figure}

 In the upper panel of Figure \ref{sigma_sr} we see results for the time dependence of the total stress $\sigma$, at a large and a small value of $k_0$, and for three values of $L$. For large $k_0$ we see the same qualitative situation as in MF: $\sigma(t)$ displays statistical fluctuations that decrease with $N$ according to the expected behavior of the standard deviation of $\sigma(t)$ decreasing as $1/L$.
For small $k_0$ we also observe a situation comparable with MF: $\sigma(t)$ displays an oscillatory behavior, and the span of the oscillation becomes system size independent when this is large. 
This is a convincing evidence that global oscillations can be observed in the case of NN interactions.
There is however a remarkable qualitative difference between the form of $\sigma(t)$ here and that in MF (Fig. \ref{detalle}(b)). In the present case and as system size increases, $\sigma(t)$ never displays an abrupt instantaneous decay, but becomes a time-continuous function as $L\to\infty$. In other words, the avalanches that produce the stress oscillation are individually much smaller than the system size when this is large. 
This can be verified by a direct evaluation of the distribution of avalanches in time, which is shown in the lower panel of Fig. \ref{sigma_sr}. Both at the large and the small value of $k_0$, the avalanche distribution is independent of system size. However, for the large $k_0$ the size distribution is uniform in time, producing a constant value of $\sigma$ in the thermodynamic limit, whereas at the low value of $k_0$
the time distribution of avalanches becomes oscillatory, with smaller avalanches appearing in stress-increasing (stick) intervals and larger ones in the stress decreasing (slip) intervals. This behavior is independent of system size. We call this regime {\em oscillatory}, to emphasize the oscillatory nature of the stress, yet to distinguish it from the full stick-slip behavior we encountered with MF interactions. To have a general picture of the transition between the smooth and the oscillatory regime, in Fig. \ref{transition_sr} we plot the standard deviation of the $\sigma(t)$ signal as a function of $k_0$,
again for different system sizes. We concentrate in the data plotted as  continuous lines
(the dashed parts of the data in Fig. \ref{transition_sr} are a system size effect, and they are discussed below).
We see that the variance of $\sigma(t)$ is compatible in the thermodynamic limit with a vanishing at some critical value $k_{0c}$.
The avalanche size distributions though (shown in Fig \ref{size_sr}), do not seem to have any strong feature when crossing $k_{0c}$.
They follow roughly a cutoff power law, with a power law exponent in between the mean field value ($\tau=3/2$), and the value $\tau\simeq 1.27$  
corresponding to two-dimensional depinning without aging. The value of $S_{max}$, calculated as $S_{max}\equiv \langle S^2\rangle/\langle S\rangle$ increases steadily as $k_0$ is reduced, as shown in the inset to Fig. \ref{size_sr}. Yet, $S_{max}$ does not display a clear tendency to diverge at any finite $k_0$, suggesting that the oscillatory regime extends to $k_0\to 0$ in the thermodynamic limit.

Notice that if the value of $k_0$ is reduced even further, a second transition to a state with even larger value of the variance of $\sigma(t)$ is observed (dotted lines in Fig. \ref{transition_sr}).
This corresponds to a transition to a  fully synchronized regime at very
 low $k_0$,
which is similar to the stick-slip regime encountered in the MF case. However, this
 transition systematically shifts toward lower values of $k_0$ as the system size increases.
 Therefore, we conjecture that the strict stick-slip dynamics—characterized by single global
 avalanches spanning the entire system—disappears in the thermodynamic limit. Consequently, the
 oscillatory regime would be the only non-uniform dynamical state that persist in that limit.

\begin{figure}
\includegraphics[width=9cm,clip=true]{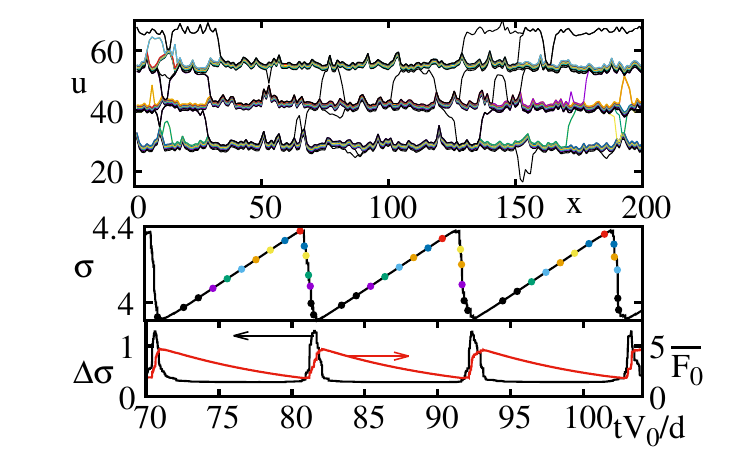}
\caption{(a) One-dimensional cuts of the interface at different times in the oscillatory regime (each successive curve has been vertically shifted a small amount for clarity),
(b) $\sigma(t)$, and (c) $\overline {F_0}(t)$ and $\Delta \sigma(t)$ curves of the system along the simulation ($k_0=0.05$, $V_0\tau/d=10$). Each cut in panel (a), from bottom to top, correlates with one point in (b), from left to right.
Note how during the ``stick" periods the interface remains pinned, and almost flat, 
whereas during ``slip", the interface transitions through a series of avalanches, all of them smaller than the system size.}
\label{temporal}
\end{figure}

The nature of the oscillatory regime we observe deserves further analysis. In order to spot in detail a temporal sequence of interface configurations, we will display one dimensional cuts of the full two-dimensional surface, at different times. In Fig. \ref{temporal} we show a sequence of configurations along time in the oscillatory regime. We observe that the interface moves between almost flat states 
that correspond qualitatively to the sticked states of the one parameter model describe in Appendix I. However the transition between two consecutive flat states does not occur through a single system-size avalanche, but in a sequence of events, none of them scaling with the system size. This process is made evident in the evolution of $\Delta\sigma(t)$ (Fig. \ref{temporal}(b)) that displays a sharp increase during a finite time interval in the slip processes.

It is important to understand what makes synchronization possible in a system with short range interactions. It turns out that a crucial role is played by the relaxation state of different parts of the sample synchronizing also when $\sigma$ displays global oscillation. In  panel (c) of Fig. \ref{temporal} we display the average value of the pinning force in the system $F_0$. We recall that for a single site, this value drops to $f_0-\beta$ after an avalanche, and exponentially recovers to $f_0$ (with time constant $\tau$) as a function of time. We see that the spatial average of $F_0$ follows qualitatively the same trend, dropping at the slip phase, and recovering during the stick periods. Notice the similarity with the behavior of the one block model (Appendix I). Yet we emphasize the difference during the slip phase, which in the extended model occurs without the appearance of system size ``king" avalanches. If from the synchronized situation of  Fig. \ref{temporal} we increase the value of $V_0$ (keeping other parameters fixed), syncronization is lost qualitatively in the same way as in the mean field case: syncronization disappears when the increase in the driving force after an avalanche (due to the velocity $V_0$) is more rapid than the strengthening of the pinning force due to relaxation. On the other hand, we have observed that syncronization is also lost when $V_0$ is reduced sufficiently. This behavior was not present in mean field, where the stick-slip phase persist up to $V_0=0$ (Fig. \ref{phase_diagram}). The reason for the lost of syncronization is related to the fact that for $V_0\to 0$, the stabilizing effect of a correlated state of relaxation across the sample disappears, as the sample fully relax after each individual avalanche.
More thorough numerical analysis would be necessary  to map the complete phase
 diagram for NN interactions, and this is left for a future work. 
 But the analysis just presented suggests that we will observe ``synchronization without kings", in a parameter region corresponding to driving with a sufficiently soft spring, and only in an intermediate range of velocities. 
 This is a highly non-trivial result with far-reaching
 consequences, which are further discussed in the next Section.

\section{Summary and Conclusions}

We have presented a model 
that gives a detailed framework for understanding the transition from standard depinning to globally synchronized states in frictional systems. The key result is that the inclusion of a local aging mechanism--the increase of the depinning force with contact time--is sufficient to generate the velocity-weakening conditions 
which act as the primary driver for global instabilities and the appearance of ``king avalanches". Through both analytical calculations  and numerical simulations, we have mapped a complex phase diagram in mean-field featuring three distinct dynamical regions: a stable stationary phase, a pure stick-slip phase, and a bistable region where the system's behavior depends on its initial conditions.


Beyond mean field,
one of the most significant findings of our research is the persistence of global synchronization states even in two-dimensional systems with short-range (nearest-neighbor) interactions. 
Our numerical simulations strongly suggest that synchronization is possible even in the thermodynamic limit ($N\to\infty$) for local interactions. However, this synchronization possesses a distinct nature compared to the mean-field case: it does not involve system-size scaling avalanches. Instead, it manifests as alternating temporal intervals of higher and lower avalanche activity that correlate with global stress oscillations. This result is crucial as it demonstrates in a concrete model that collective oscillatory behavior can emerge from purely local interaction rules.

The relevance of this study extends to a wide range of out-of-equilibrium systems that respond through intermittent bursts of activity. Application to seismic fault dynamics has been our main target as in that context velocity-weakening conditions are essential for realistically describing earthquakes and the stick-slip stress cycles.
Yet the model may find application in other fields, as in the deformation of amorphous solids
with aging, which would require to consider non positive definite generalized elastic interaction kernels of the Eshelby type.
It is also interesting to indicate potential applications in fields away of material science.
Specifically, this model provides a robust framework to address  the synchronization of neural systems. The depinning model with aging shares a profound analogy with the firing dynamics of neuronal populations. In this context, the ``aging" mechanism can be interpreted as synaptic fatigue or recovery processes, while the local pinning represents the firing threshold. The model's ability to exhibit synchronized activity intervals (bursting) under short-range interactions provides a theoretical basis for how large brain regions can coordinate electrical activity without total connectivity. This is particularly relevant for understanding both cognitive processes and pathological states characterized by abnormal synchronization.

In conclusion, the inclusion of aging in depinning models not only enriches the physics of dynamic phase transitions but also provides a versatile framework for studying synchronization in natural systems where interactions are inherently local and history-dependent.

\section{Acknowledgments}

We kindly acknowledge stimulating discussions with Alberto Rosso, Eugenio Lipiello and Giuseppe Petrillo.

\section{Appendix I}
\label{ap1}

\begin{figure}[h!]
    \centering
    \includegraphics[width=0.8\linewidth]{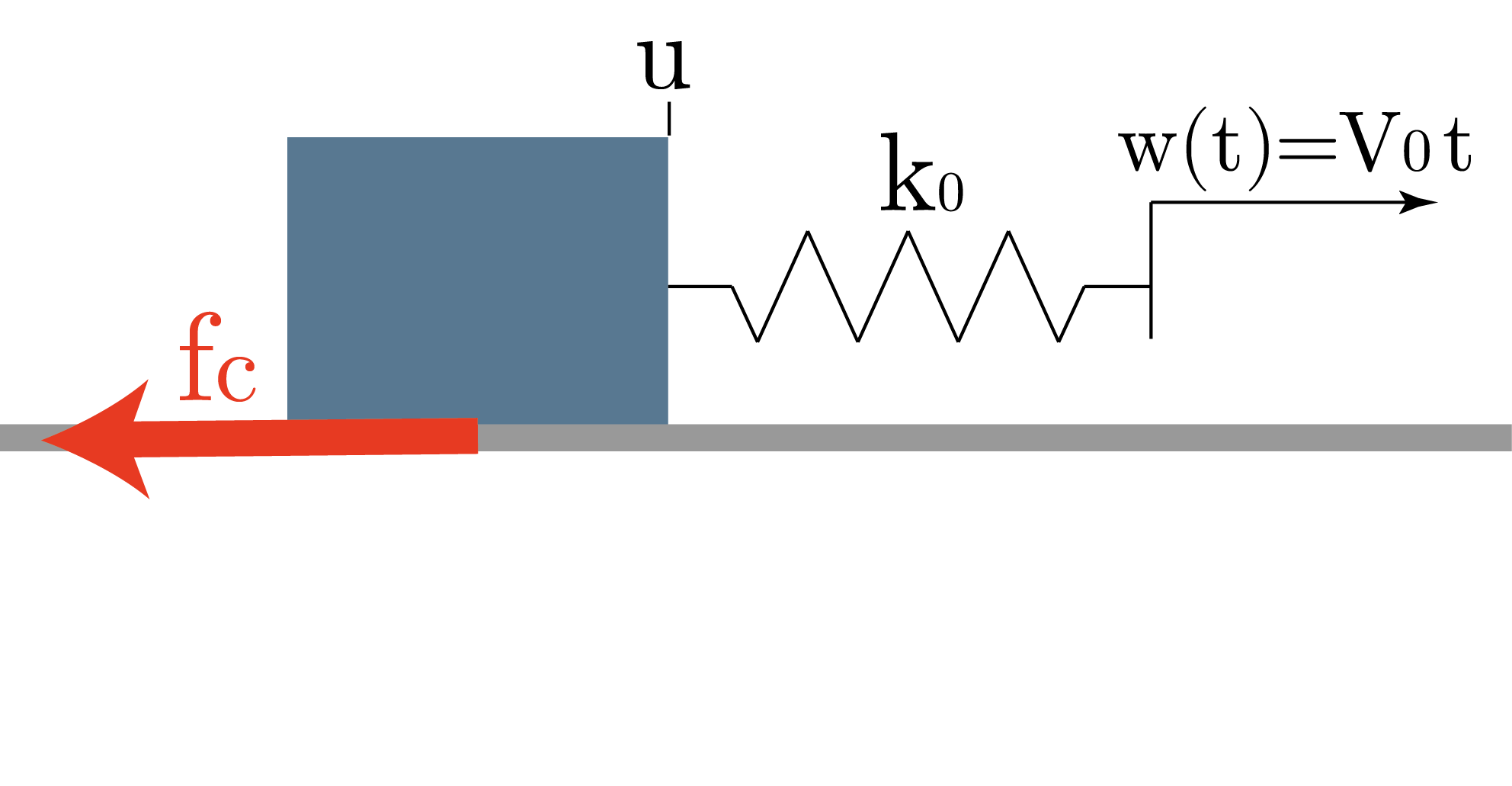}
    \caption[Block–spring system schematic]{Schematic diagram of the system considered: a rigid block driven by a spring of stiffness $k_0$, sliding over a surface where friction is given by $f_c$.
    }
    \label{fig:ap2_bloque_resorte}
\end{figure}

Let us consider a system such as the one depicted in Fig.~\ref{fig:ap2_bloque_resorte}, consisting of a single rigid block sliding over a fixed substrate. The block is driven by a spring of stiffness $k_0$, whose opposite end, at coordinate $w$, moves at constant velocity, $w = V_0 t$. The position of the block is denoted by $u$, and its velocity by $v = \dot{u}$. The spring exerts an elastic force $f_k = k_0 (w - u)$.

In addition to the spring force, the block is subjected to a friction force $f_c$. This force depends on whether the block is moving or stationary and, in the latter case, on the time the block has remained at rest (aging). The friction force is defined as

\begin{equation}
f_c(v,T) = 
\begin{cases}
f_0 - \beta, & \text{if } v > 0, \\
f_0 - \beta \exp(-T/\tau), & \text{if } v = 0,
\end{cases}
\label{fc}
\end{equation}
where $\tau$ is the characteristic aging time, and $T$ denotes the time elapsed since the block stopped moving. The dynamics of the system is determined by the condition that the net force on the block vanishes. We explicitly assume the block has zero mass, meaning inertial effects are neglected.

Under these conditions, suppose the block initially remains at rest for a sufficiently long time at some position $u_0$, implying $f_c = f_0$. When the spring begins to stretch at constant velocity $V_0$, the elastic force increases as $f_k = k_0 V_0 t$. The spring continues to load until the elastic force equals the friction force, at which point the block begins to move. At that instant, the friction force drops to $f_0 - \beta$, causing the block to slip to a position where $f_k = f_0 - \beta$. 

From this moment on, both forces begin to increase again: $f_k$ grows linearly in time, while $f_c$ evolves according to Eq.~\eqref{fc}. Two situations may then arise:

1: The recovery of the friction force is faster than the increase of the elastic force, or  
2: The increase of the elastic force is faster than the recovery of the friction force.

\begin{figure}[h!]
    \centering
    \includegraphics[width=0.8\linewidth]{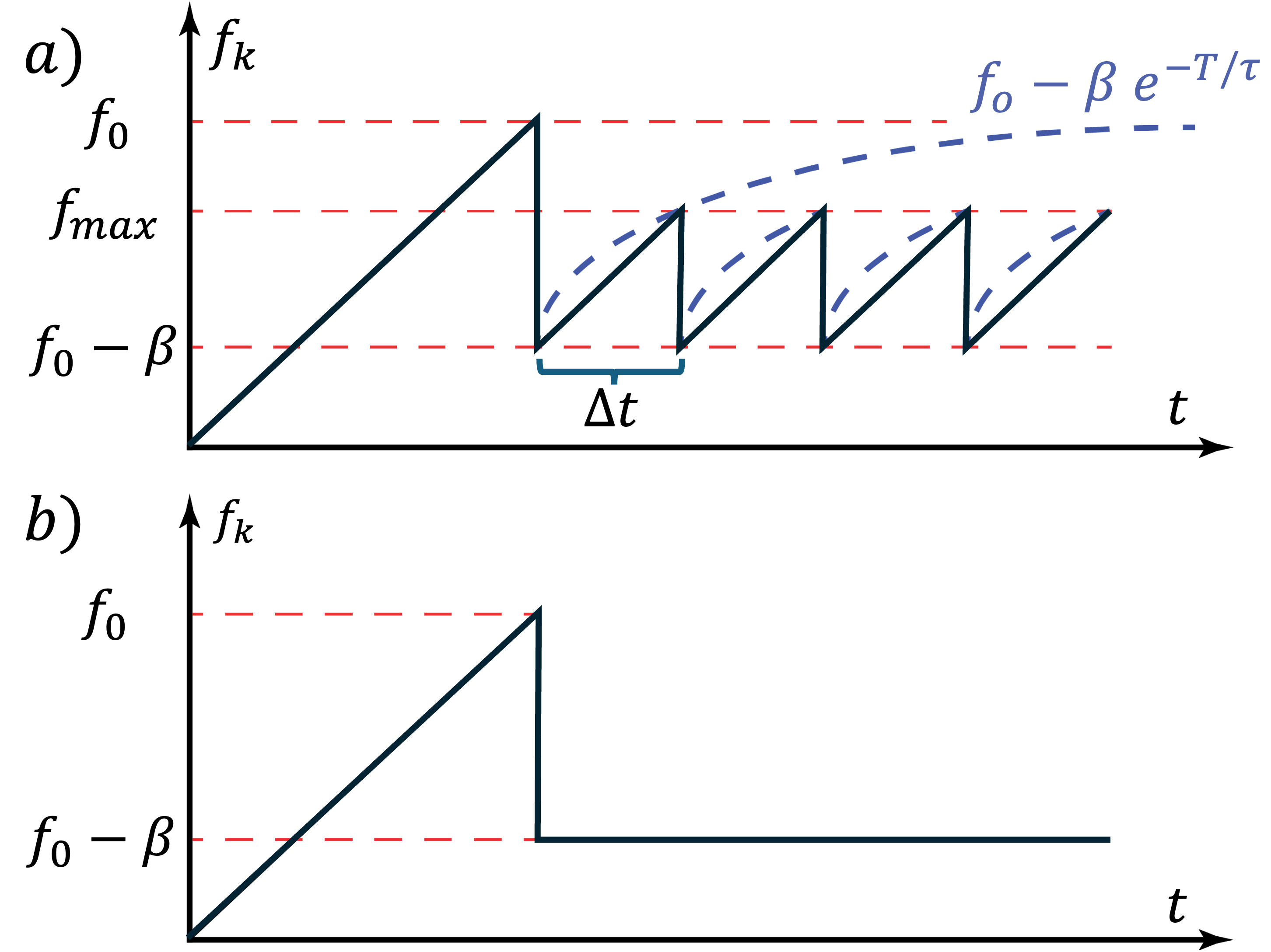}
    \caption[Schematic evolution of the spring force in the block--spring system]{Schematic evolution of the spring force $f_k$ for cases 1 and 2 described in the text. Case 1 corresponds to stick-slip dynamics, whereas case 2 exhibits smooth sliding.}
    \label{fig:ap2_sigmas}
\end{figure}

In case 1, the block remains stationary until the two curves intersect, which occurs when

\begin{equation}
\beta \left(1 - e^{-\Delta t / \tau}\right) = k_0 V_0 \Delta t.
\label{cruce}
\end{equation}
Under these conditions, the dynamics is of stick-slip type, with the extreme values of the elastic force given by  
$f_k^{\min} = f_0 - \beta$ and  
$f_k^{\max} = f_0 - \beta \exp(-\Delta t/\tau)$.  
Figure~\ref{fig:ap2_sigmas}(a) illustrates this situation.

Conversely, in case 2, the elastic force increases more rapidly than the rate of friction force recovering, resulting in the block being continuously in motion. The force balance is then simply  
$f_k = f_0 - \beta$.  
This scenario is illustrated in Fig.~\ref{fig:ap2_sigmas}(b).

The transition between these two regimes occurs when, immediately after motion begins, the time derivatives of the spring force and the recovering friction force become equal. This yields the critical condition separating both behaviors:

\begin{equation}
\frac{\beta}{\tau} = k_0 V_0.
\label{lineac}
\end{equation}
Equation~\eqref{lineac} defines a boundary in the $(k_0, V_0)$ parameter plane: the upper-right region corresponds to smooth sliding, whereas the lower-left region corresponds to stick-slip dynamics. Figure~\ref{phase_diagram} shows the transition line obtained from the analysis presented in this appendix, superimposed on the phase diagram extracted from the numerical simulations of the model developed in this work. 

Note that when crossing this transition line, the oscillation amplitude changes from zero in the smooth phase to a nonzero value $\delta$ in the stick-slip phase, where $\delta$ grows as the square root of the distance to the transition line. In other words, the qualitative behavior obtained here mirrors the phase-diagram structure found in the mean-field analysis for large values of $V_0$ (Fig.~\ref{phase_diagram}).

\section{Appendix II}
\label{ap2}

We give here the derivation of the size distribution of avalanches for mean field interactions. We first analyze the case in which there are no aging effects ($\beta=0$), and afterwards we indicate the modifications that appear when $\beta>0$. 

\begin{figure}
\includegraphics[width=6cm,clip=true]{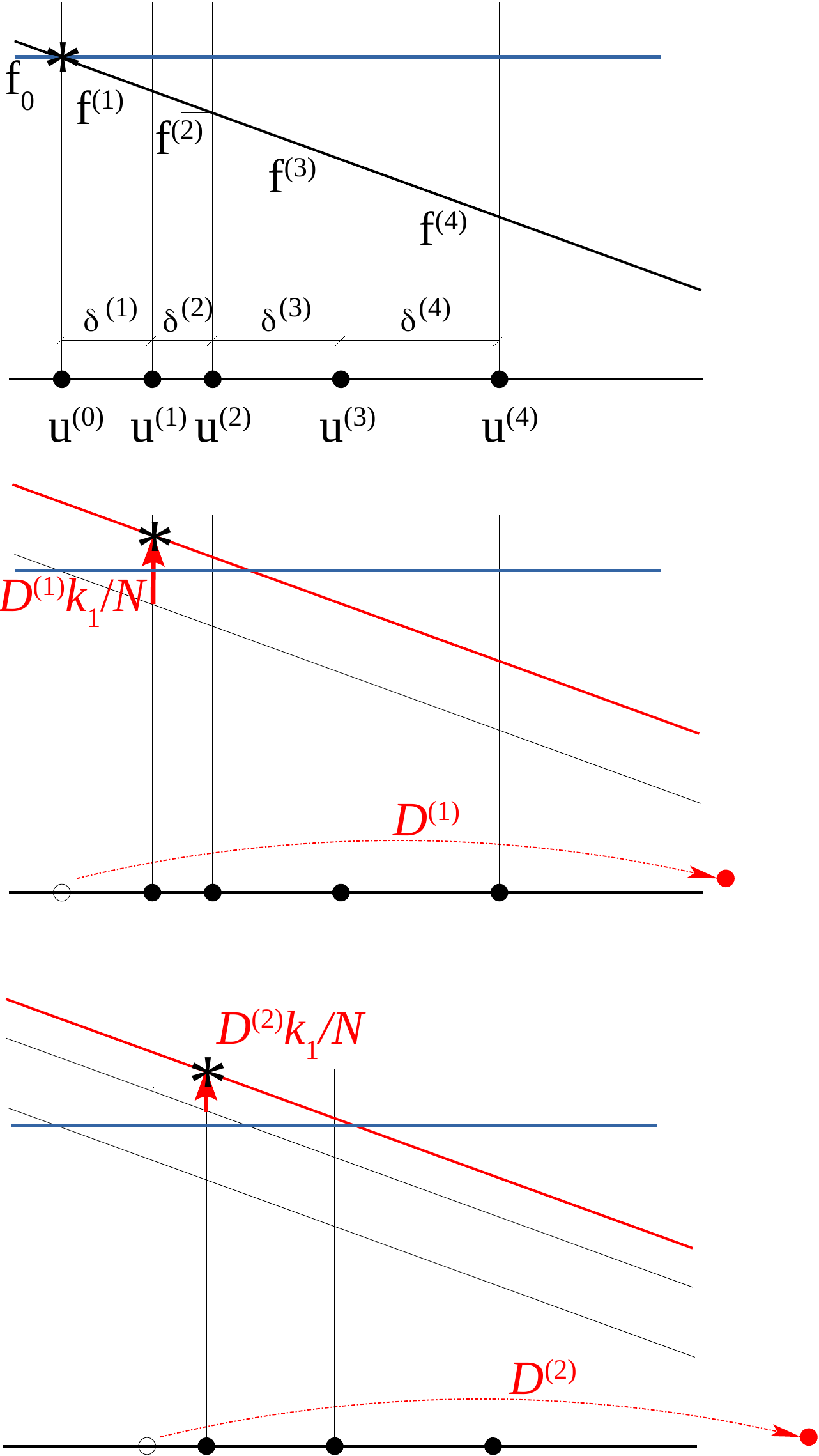}
\caption{Sketch of the development of an avalanche in mean field. The instability of the first site $u^{(0)}$ may cause a cascade of additional instabilities. The process continues after $j$ steps if
$(k_0+k_1)\sum_{i=1}^j\delta^{(i)}>(k_1\sum_{i=1}^j D^{(i)})/N$.
}
\label{app}
\end{figure}

For the $\beta=0$ case, the starting point is the equilibrium distribution ${\cal P}(u)$ that we reproduced here to facilitate the 
analysis
\begin{eqnarray}
{\cal P}(u)&=&d^{-1}\exp(-(u-u_0)/d) ~~~ \mbox {if} ~~~u>u_0\\
{\cal P}(u)&=&0 ~~~~~~~~~~~~~~~~~~\mbox {if} ~~~u<u_0 
\label{pofu}
\end{eqnarray}
with
$u_0=w-(f_0-k_1d)/k_0$.
In order to calculate the size of the next avalanche in the system, we must look in detail at the actual values of the variables $u_i$. It will be convenient to consider them sorted in increasing order by introducing an upper-script:
\begin{equation}
u^{(0)}<u^{(1)}<u^{(2)}<...
\end{equation}
Upon an increase of the driving $w$, $u^{(0)}$ will be the variable that destabilizes first, i.e., the force onto it reaches the critical value $f_0$. At this point, (and without increasing further the value of $w$) we have to see how many sites become unstable and cascade to generate the full avalanche.
The process can be described in the following terms (see Fig. \ref{app}).
When $u^{(0)}$ becomes unstable, it jumps to the next potential well which is at a position 
$u^{(0)}+D^{(1)}$ where $D^{(1)}$ is taken from a Poisson distribution with mean $d$. This increase in $u^{(0)}$ produces a change of $\overline u$, and therefore an increase in the force on all other sites given by $\Delta f^{(i)}=D^{(1)}k_1/N$. We have to evaluate now if this force increment is able to destabilize $u^{(1)}$, namely if the condition
\begin{equation}
f^{(1)}+\frac{D^{(1)}}{N}k_1>f_0
\end{equation}
is satisfied. If this is the case, $u^{(1)}$ will jump to a new position $u^{(1)}+D^{(2)}$, and the process is repeated. We now have to evaluate if 
\begin{equation}
f^{(2)}+\frac{D^{(1)}}{N}k_1+\frac{D^{(2)}}{N}k_1>f_0
\end{equation}
to see if the avalanche goes on $u^{(2)}$.
In general terms, we continue the avalanche until the lowest $j$ for which 
\begin{equation}
f^{(j)}+\sum_{i=1}^j \frac{D^{(i)}}{N}k_1<f_0.
\label{fj}
\end{equation}
When this first happens, an avalanche of size 
\begin{equation}
S=\sum_{i=1}^{j-1} D^{(j)}
\label{s}
\end{equation}
 has occurred, and the system reaches a stable configuration. 
Referring to Fig. \ref{app}, calling $\delta^{(i)}=u^{(i)}-u^{(i-1)}$, we see that we can write 
\begin{equation}
f^{(j)}=f_0-(k_0+k_1)\sum_{i=1}^j \delta^{(i)}
\end{equation}
in such a way that the condition for the avalanche to stop (Eq. (\ref{fj})) can be written as
\begin{equation}
\sum_{i=1}^j \left (k_1\frac{D^{(i)}}{N}-(k_0+k_1)\delta^{(i)}\right )<0.
\end{equation}
As the values of $u^{(i)}$ are uncorrelated, the separation variables $\delta^{(i)}$ have Poissonian distribution with mean value $d/N$, exactly as the variable $D^{(i)}/{N}$. Therefore,  defining
 $\widetilde \delta^{(i)}\equiv D^{(i)}/N$,  
  we can define an effective random walk with step 
\begin{equation}
\eta_i\equiv \left (k_1{\widetilde \delta^{(i)}}-(k_0+k_1)\delta^{(i)}\right ).
\label{esta}
\end{equation}
The  avalanche stops at the first $j$ for which
\begin{equation}
\sum_{i=1}^j \eta^{(i)} < 0
\end{equation}
Therefore, the stop of the avalanche corresponds to the first instance in which the RW (which starts at 0) returns back to the origin, i.e., we can identify the size of the avalanche as proportional to the  number of discrete steps it takes to the RW to return to 0 for the first time. Notice that the RW $\eta^{(i)}$ is biased: $\langle \eta\rangle =-k_0d$. In addition, its variance 
is $\sigma_\eta^2=d^2(k_0+k_1)k_1$.


Focusing in the distribution of large  avalanches, note that if they involve many unstable sites we can write Eq. (\ref{s}) as $S\simeq d j$, where $j$ is the number of steps of the random work at the first return to the origin, and $d$ the average separation between pinning wells. 
Analytical expressions can be derive by  analyzing this RW in the continuum limit. 
We consider the distribution function $W(x,t)$ of RWs $\eta_i$ with the previously given values of $\langle \eta \rangle$ and $\sigma_\eta$. The Fokker-Planck equation satisfied by $W(x,t)$ is a diffusion equation with bias:
\begin{equation}
\frac{\partial W}{\partial t}=\frac{\sigma_\eta^2}{2}\frac{\partial^2 W}{\partial x^2} +\langle\eta\rangle \frac{\partial W}{\partial x}
\label{fp}
\end{equation}
(here $t$ is non-dimensional, and proportional to the number of steps $i$).
The boundary/initial conditions that have to be imposed on $W$ to represent our RWs are:
$W(x>0,t=0)=0$ (the RWs start at $x=0$ at $t=0$), and $W(x=0,t)=0$ (RWs annihilate when reaching $x=0$ for the first time).
The survival probability of  a RW at time $t$ is given by $\int_0^\infty W(x,t) dx$, and  the probability that the RW reaches the origin exactly at time $t$  is given by 
\begin{equation}
P(t)=\frac d{dt }\int_0^\infty W(x,t) dx
\end{equation}
By integrating Eq. (\ref{fp}) in $x$ between 0 and $\infty$, we can write this expression as
\begin{equation}
P(t)\sim \left .\frac {d W(x,t)}{dx}\right |_{x=0}
\label{borde}
\end{equation}

We are left with the calculation of $W(x,t)$. 
Let us first consider the case $\langle\eta\rangle=0$. We can construct the appropriate solution in this case using a diffusive Gaussian  pulse to solve the diffusion equation, and taking an $x$-derivative which produces the appropriate boundary condition at $x=0$.
Namely
\begin{eqnarray}
W(x,t)&\sim&\frac{d}{dx} \left (t^{-1/2}e^{-\frac{x^2}{2\sigma_\eta^2 t}}\right )\\
&\sim&{x}{t^{-3/2}}e^{-\frac{x^2}{2\sigma_\eta^2 t}}.
\end{eqnarray}
Then, using Eq. (\ref{borde}), identifying time with the avalanche size $S$ we obtain
\begin{equation}
P(S)\sim S^{-3/2}
\end{equation}

Now let us consider the case $\langle  \eta\rangle \ne 0$.
The natural starting solution to Eq. (\ref{fp}) is  now
\begin{equation}
W_0(x,t)=t^{-1/2}e^{-\frac{(x-\langle\eta\rangle t)^2}{2\sigma_\eta^2 t}} 
\end{equation}
However, in the present case, the function $W_1(x,t)=dW_0(x,t)/dx$ does not satisfy the condition that it vanishes at $x=0$. However, we can construct a function that vanishes at $x=0$ and is still a solution of Eq. (\ref{fp}) by making a linear combination of $W_0$ and $W_1$. In fact, the function 
\begin{equation}
W(x,t)\equiv\frac{\langle \eta\rangle}{\sigma_\eta^2}W_0(x,t)- W_1 (x,t)
\end{equation}
is still solution of Eq. (\ref{fp}), and satisfies $W(0,t)=0$.
Using Eq. (\ref{borde}) and the identification $S\simeq td$ we obtain the size distribution of avalanches in this case as
\begin{equation}
P(S)\sim S^{-3/2}e^{-\frac{{\langle \eta\rangle}^2S}{2\sigma^2_\eta d}}
\end{equation}
Using the values of $\langle \eta\rangle$ and $\sigma_\eta$ appropriate for our case we obtain finally
\begin{equation}
{\cal P}(S)\sim S^{-3/2}e^{-S/S_{max}}
\end{equation}
with ${S_{max}=2d(k_0+k_1)k_1/{k_0^2}}$. This value is finite for any $k_0$, diverging only when $k_0 \to 0$, as $S_{max}\sim k_0^{-2}$.

In the case in which there is aging in the system  ($\beta>0$), the calculation proceeds along similar lines, with the main difference that instead of the form of ${\cal P}(u)$ as given by Eq. (\ref{pofu}), we have to use the solution appropriate to this case given by (\ref{eq:dist_general}). The calculation is straightforward but a bit involved.
We summarize the main results that are obtained. When interested in large avalanches, the problem can still be mapped to the problem of the first return to zero of a biased random walk $\eta$. 
This at once tells that the form $P(S)\sim S^{-3/2}\exp{(-S/S_{max})}$ is maintained. 
The qualitative difference with the previous case is that the bias of this random walk, i.e., the value of $\langle \eta\rangle$ differs from the expression obtained for $\beta=0$. 
In that case we had $\langle \eta\rangle=-k_0 d$, and its negative value is  the indication that the random walk returns to zero with probability 1 in a finite time, which in turn means that all avalanches have finite size. 
For an infinitely large system this means that the stress fluctuation caused by any avalanche is vanishingly small, which confirms the assumed constant smooth dynamics with a constant value of stress.
The calculation of $\langle \eta\rangle$ in the case $\beta>0$ shows parameter regions where $\langle \eta\rangle<0$ and others where $\langle \eta\rangle>0$. In the first case this corresponds to a stable smooth dynamics. However when $\langle \eta\rangle>0$ the conclusion is that there is a finite probability that the random walk never returns to the origin. Speaking in terms of avalanches, this means that there is a finite probability that an avalanche never stops, or in other words, that it reaches a size comparable to system size, no matter how large this is. If this happens, the assumption that the dynamics is smooth cannot be maintained. We can therefore use the limiting case $\langle \eta\rangle=0$ to obtain the border of the region for which a smooth dynamics is possible. The result (which is obtained in the next section) is incorporated into the phase diagram presented in Fig. \ref{phase_diagram}.

\subsection{Calculation of the value of $\langle \eta\rangle$ in the presence of aging}

In the presence of aging, the definition of an effective random walk in the form  (see Eq. (\ref{esta}))
\begin{equation}
\eta^{(i)}=k_1 {\widetilde\delta^{(i)}}-(k_0+k_1) \delta^{(i)}.
\end{equation}
is still valid.
The first term is the force increase on every site 
given that some other site
$i$ has jumped a distance $D^{(i)}=N\widetilde\delta^{(i)}$ from a potential well to the next stable well. In the second term $\delta^{(i)}$  is the separation between successive $u$-values of sites that are close to the instability threshold (see Fig. \ref{app}). 
When the two terms are equal on average,  the bias of the effective random walk is zero, and  we are in a situation in which an infinite size avalanche can occur, therefore marking the limit to a possible smooth dynamical behavior in the system. 
In the absence of relaxation we have $\overline \delta=\overline {\widetilde \delta}=d/N$, and $\langle \eta\rangle$ becomes zero only if $k_0\to 0$. 
In the presence of relaxation the values of $\overline \delta$ and $\overline {\widetilde \delta}$ change, and must be evaluated with care in terms of the distribution ${\cal P}(u,T)$ (Eq. (\ref{pdeut})). This is done as follows.
In the presence of relaxation if a particle that has spent a time $T$ on a given well is destabilized, the average jump distance $\overline D$ is larger than the average inter-well distance $d$. This is because the force 
at depinning from a given well occurs at a value of force $F_0(t)=f_0-\beta\exp(-T/\tau)$ that is larger than the pinning force of the new well, which is always $f_0-\beta$. The particle is not pinned in the new well if the separation between the two is smaller than $A_0(1-\exp(-T/\tau))$ (where $A_0\equiv \beta/(k_0+k_1)$). If this happens, the particle jumps immediately to the next well.
This additional separation must be added to the average value $d$ between wells to calculate the average jump distance of a particle, which results
\begin{equation}
\overline D=N\overline{\widetilde\delta}=d+A_0\left (1-\overline{\exp(-T/\tau)}\right )
\end{equation}
The average sign over the exponential factor indicates that we have to average over the different $T$ values of particle that could potentially be destabilized. This average is evaluated in terms of the known distribution ${\cal P}(u,T)$ (Eq. (\ref{pdeut})) as
\begin{equation}
\overline{\exp(-T/\tau)}=\int^\infty_{T^*} dT P(U_0(T),T)\exp(-T/\tau).
\end{equation}
Notice that the lower integration limit is not directly set to 0, since in the case $V_0<d/\tau$ there are no particles at the instability point if $0< T< T^*$  (see Fig. \ref{put}). This means that we have to take $T^*=0$ only if $V_0>d/\tau$, but in the case $V_0<d/\tau$ we have to choose $T^*$ as the positive solution of the equation
\begin{equation}
A_0\exp(-T^*/\tau)=A_0-V_0T^*.
\end{equation}

Now we have to calculate $\overline {\delta}$ which, as we indicated before, is the average separation between $u$ values of particles that are close to destabilization. Referring again to Fig. \ref{app}, we can calculate the density of particles at the destabilization threshold as
\begin{equation}
\int^\infty_{T^*} dT P(U_0(T),T)
\end{equation}
and from here 
\begin{equation}
\overline {\delta}=\left (\int^\infty_{T^*} dT P(U_0(T),T)\right )^{-1}
\end{equation}

In this way, the numerical values of $\overline {\widetilde \delta}$ and $\overline {\delta}$ can be calculated (using numerical integration) using the expressions we have obtained, and the known form of $P(u,T)$ (Eq. (\ref{pdeut})). As long as 
 $k_1\overline {\widetilde\delta}>(k_0+k_1)\overline {\delta}$ the random walk representing the process has a negative bias, and produces a finite value of $S_{max}$. When $k_1\overline {\widetilde\delta}=(k_0+k_1)\overline {\delta}$ the bias is zero, and $S_{max}$ diverges, indicating the stability limit of the smooth dynamical phase has been reached. These points are indicated in the phase diagram of Fig. \ref{phase_diagram} by the continuous line, which very nicely agrees with and reinforces the results of the numerical simulations. Notice that the abrupt change of direction of this line at some particular value of $V_0$ separates the case where $T^*=0$ (upper part of the curve)
 from that where $T^*>0$ (lower part of the curve).

\bibliography{bib}

\end{document}